\documentclass[fleqn,usenatbib]{mnras}

\usepackage[T1]{fontenc}
\usepackage{ae,aecompl}

\usepackage{graphicx}	% Including figure files
\usepackage{amsmath}	% Advanced maths commands
\usepackage{amssymb}	% Extra maths symbols
\usepackage{longtable} 
\usepackage{hyperref}
\usepackage{cleveref}

\newcommand{\gaia}{{\it Gaia}}
\newcommand{\ha}{H$\alpha$}
\title[XGAPS]{XGAPS: a sub-arcsecond cross-match of Galactic Plane Surveys}

\author[S. Scaringi et al.]{S. Scaringi$^{1}$\thanks{e-mail: simone.scaringi@durham.ac.uk}, 
M. Mongui\'o$^{2,3,4}$,
C. Knigge$^{5}$,
M. Fratta$^{1}$,
B. G{\"a}nsicke$^{6}$,
\newauthor
P. J. Groot$^{7,8,9}$,
A. Rebassa-Mansergas$^{10,4}$,
O. Toloza$^{11}$.
\\
$^{1}$Centre for Extragalactic Astronomy, Department of Physics, Durham University, South Road, Durham, DH1 3LE\\
$^{2}$Institut de Ci\`encies del Cosmos (ICCUB), Universitat de Barcelona (UB), Mart\'i i Franqu\`es 1, E-08028 Barcelona, Spain\\
$^{3}$Departament de Física Qu\`antica i Astrof\'isica (FQA), Universitat de Barcelona (UB), Mart\'i i Franqu\`es 1, E-08028 Barcelona, Spain\\
$^{4}$Institut d'Estudis Espacials de Catalunya (IEEC), c. Gran Capit\`a, 2-4, 08034 Barcelona, Spain\\
$^{5}$School of Physics and Astronomy, University of Southampton, Highfield, Southampton SO17 1BJ, UK\\
$^{6}$Astronomy and Astrophysics Group, Department of Physics, University of Warwick, Gibbet Hill Road, Coventry, CV4 7AL, UK\\
$^{7}$Department of Astrophysics/IMAPP, Radboud University, P.O. 9010, 6500 GL, Nijmegen, The Netherlands\\
$^{8}$South African Astronomical Observatory, PO Box 9, Observatory, 7935, Cape Town, South Africa\\
$^{9}$Department of Astronomy, University of Cape Town, Private Bag X3, Rondebosch, 7701, South Africa\\
$^{10}$Departament de F\'isica, Universitat Polit\`ecnica de Catalunya, c/Esteve Terrades 5, 08860 Castelldefels, Spain\\
$^{11}$Departamento de F{\'i}sica, Universidad T{\'e}cnica Federico Santa Mar\'ia, A. Espa{\~n}a 1680, Valpara\'iso, Chile\\
}

\date{Accepted 14 November 2022. Received in original form 14 September 2022.}

\pubyear{2022}

\begin{document}
\label{firstpage}
\pagerange{\pageref{firstpage}--\pageref{lastpage}}
\maketitle

\begin{abstract}
We present a sub-arcsecond cross-match of \gaia\ Data Release 3 (DR3) against the INT Galactic Plane Surveys (IGAPS) and the United Kingdom Infrared Deep Sky Survey (UKIDSS). The resulting cross-match of Galactic Plane Surveys (XGAPS) provides additional precise photometry ($U_{RGO}$, $g$, $r$, $i$, H$\alpha$, $J$, $H$ and $K$) to the \gaia\ photometry. In building the catalogue, proper motions given in \gaia\ DR3 are wound back to match the epochs of the IGAPS constituent surveys (INT Photometric \ha Survey of the Northern Galactic Plane, IPHAS, and the UV-Excess Survey of the northern Galactic plane, UVEX) and UKIDSS, ensuring high proper motion objects are appropriately cross-matched. The catalogue contains 33,987,180 sources. The requirement of $>3\sigma$ parallax detection for every included source means that distances out to 1--1.5 kpc are well covered. In producing XGAPS we have also trained a Random Forest classifier to discern targets with problematic astrometric solutions. Selection cuts based on the classifier results can be used to clean colour-magnitude and colour-colour diagrams in a controlled and justified manner, as well as producing subsets of astrometrically reliable targets. We provide XGAPS as a 111 column table. Uses of the catalogue include the selection of Galactic targets for multi-object spectroscopic surveys as well as identification of specific Galactic populations.
\end{abstract}

\begin{keywords}
catalogues -- surveys -- parallaxes -- proper motions -- stars:emission-line -- Galaxy: stellar content
\end{keywords}

%%%%%%%%%%%%%%%%%%%%%%%%%%%%%%%%%%%%%%%%%%%%%%%%%%

%%%%%%%%%%%%%%%%% BODY OF PAPER %%%%%%%%%%%%%%%%%%

\section{Introduction}

The European Space Agency's \gaia\ mission (\citealt{gaiaMission}) Early Data Release 3 (EDR3) provides photometry in the $G$, $G_{BP}$ and $G_{RP}$ bands, as well as precise astrometry and parallax measurements for over 1.5 billion sources (\citealt{gaiaEDR3,gaiaEDR3Astro,gaiaEDR3Phot,gaiaEDR3Plx}). Although the absolute number of sources is comparable to \gaia\ Data Release 2 (DR2; \citealt{gaiaDR2,gaiaDR2Astro,gaiaDR2Cat,gaiaDR2Phot,gaiaDR2Plx}), the astrometric and photometric precision has drastically improved thanks to a 3-fold increase in the celestial reference sources and longer data collection baseline (22 vs 34 months), as well as an updated and improved processing pipeline (\citealt{gaiaEDR3Astro}). This quantity and quality is defining a new standard for Galactic studies. The more recent \gaia\ Data Release 3 (DR3) augments EDR3 by providing additional information on some detected targets such as variability indicators, radial velocity, binary star information, as well as low-resolution spectra for >200 million sources (e.g. \citealt{gaiaDR3cat,gaiaDR3var,gaiaDR3rv,gaiaDR3gold,deangeli22}). 

The INT/WFC Photometric H$\alpha$ Survey of the Northern Galactic Plane (IPHAS; \citealt{iphas}) is the first comprehensive digital survey of the northern Galactic disc ($|b|<5^\circ$), covering a Galactic longitude range of $29^\circ < l < 215^\circ$. The IPHAS observations are obtained using the Wide Field Camera (WFC) at the prime focus of the 2.5m Isaac Newton Telescope (INT) on La Palma, Spain. IPHAS images are taken through three filters: a narrow-band H$\alpha$, and two broad-band Sloan $r$ and $i$ filters. The UV-Excess Survey of the northern Galactic Plane (UVEX; \citealt{uvex}) has covered the same footprint as IPHAS using the same WFC on the INT in the two broad-band Sloan $r$ and $g$ filters as well as a Sloan $u$-like $U_{RGO}$ filter. Exposures are set to reach an $r$-band depth of $\approx 21$ in both surveys. Pipeline data reduction for both surveys is handled by the Cambridge Astronomical Survey Unit (CASU). Further details on the data acquisition and pipeline reduction can be found in \cite{iphas}, \cite{uvex} and \cite{iphasEDR}. A defining feature of both these surveys are the quasi-contemporaneous observations of each filter set so as to recover reliable colour information for sources without the contributing effects of variability on timescales longer than $\approx 10$ minutes. This same characteristic is also shared by the \gaia\ mission. Recently \cite{igaps} has produced the IGAPS merged catalogue of IPHAS and UVEX observations, while \cite{greimel21} provides the IGAPS images. Additionally to merging the sources observed by both IPHAS and UVEX, a global photometric calibration has been performed on IGAPS, which resulted in photometry being internally reproducible to 0.02 magnitudes (up to magnitudes of $\approx 18-19$, depending on the band) for all except the $U_{RGO}$ band. Furthermore, this 174-column catalogue provides astrometry for both the IPHAS and UVEX observations as well as the observation epoch, which allows to perform a precise cross-match with \gaia\ given the proper motion information provided. The astrometric solution of IGAPS is based on \gaia\ DR2. Although no per source errors are available, the astrometric solution yields typical astrometric errors in the $r$ band of 38mas.

The United Kingdom Infrared Deep Sky Survey (UKIDSS; \citealt{ukidss}) is composed of five public surveys of varying depth and area coverage which began in May 2005. UKIDSS uses the Wide Field Camera (WFCAM, see \citealt{casali07}) on the United Kingdom Infrared Telescope (UKIRT). All data is reduced and calibrated at the Cambridge Astronomical Survey Unit (CASU) using a dedicated software pipeline and are then transferred to the WFCAM Science Archive (WSA; \citealt{hambly08}) in Edinburgh. There, the data are ingested and detections in the different passbands are merged. The UKIDSS Galactic Plane Survey (GPS; \citealt{ukidssGPS}) is one of the five UKIDSS public surveys. UKIDSS GPS covers most of the northern Galactic plane in the $J$, $H$ and $K$ filters for objects with declination less than 60 degrees, and contains in excess of a billion sources. We use in this work UKIDSS/GPS Data Release 11 (DR11). Similarly to IGAPS, no per source errors are available, but the astrometric solution of UKIDSS based on \gaia\ DR2 yields a typical astrometric error of 90 mas.   
 
\cite{gIPHAS} described and provided a sub-arcsecond cross-match of \gaia\ DR2 against IPHAS. The resulting value-added catalogue provided additional precise photometry for close to 8 million sources in the northern Galactic plane in the $r$, $i$, and \ha\ bands. This paper describes a sub-arcsecond cross-match between \gaia/DR3, IGAPS and UKIDSS GPS. Similarly to \cite{gIPHAS} this cross-match of northern Galactic plane surveys (XGAPS) takes into account the different epochs of observations of all surveys and the \gaia\ astrometric information (including proper motions) to achieve sub-arcsecond precision when cross-matching the various surveys. XGAPS provides photometry in up to 9 photometric bands ($U$, $g$, $r$, $i$, \ha, $J$, $H$, $K$, $BP$, $RP$ and $G$) for 33,987,180 sources. XGAPS also provides a quality flag indicating the reliability of the \gaia\ astrometric solution for each source, which has been inferred through the use of Random Forests (\citealt{RF}). Section \ref{sec:crossMatch} describes our cross-matching procedure, including the preliminary selection cuts applied to all datasets. Section \ref{sec:RF} describes the machine learning model (using Random Forests) to train and select sources from the XGAPS catalogue which can be considered to have reliable \gaia\ astrometry, while Section \ref{sec:catalogue} describes a potential application for selecting blue-excess sources for spectroscopic follow-up. Finally conclusions are drawn in Section \ref{sec:conclusion}, and the catalogue format is summarised in the appendix.

\section{Cross-matching Gaia with IGAPS and UKIDSS} \label{sec:crossMatch}

The aim of XGAPS is to cross-match all sources detected in IGAPS (either IPHAS or UVEX) to \gaia\ DR3, and as a second step cross-match those sources to UKIDSS. The cross match is restricted to sources with a significant \gaia\ DR3 parallax detection and IGAPS sources identified as being stellar-like. 

\subsection{Selection cuts} \label{sec:selection}
Before the cross-match some selection cuts are applied to the master catalogues.

From \gaia\ DR3 only objects satisfying the following are selected:
\begin{itemize}
\item Are within an area slightly larger than the IGAPS footprint ($20<l<220$ and $-6<b<6$)
\item Have a signal-to-noise $G$-band detection above 3 (\texttt{phot\_g\_mean\_flux\_over\_error}>3);
\item Have a signal-to-noise parallax measurement above 3 (\texttt{parallax\_over\_error}>3).
\end{itemize}
\noindent This results in 41,572,231 sources. For reference, the removal of the two signal-to-noise limits would result in 240,725,104 \gaia\ DR3 sources within the IGAPS footprint. The parallax signal-to-noise limit ensures distances up to 1--1.5 kpc are well covered.

Because IGAPS is already a merge between IPHAS and UVEX, the selection cuts are applied to the individual surveys. For IPHAS detections, sources are retained only if the $r$, $i$ and \ha\ detections are not flagged as either saturated, vignetted, contaminated by bad pixels, flagged as noise-like, or truncated at the edge of the CCD. For UVEX the same cut as IPHAS is applied to the $U$, $g$, $r$ detections with the additional constraint that detections are not located in the degraded area of the $g$-band filter. Of the 295.4 million sources in IGAPS, 212,378,160 are retained through the IPHAS selection cuts and 221,495,812 are retained through the UVEX ones.

Finally the UKIDSS Galactic Plane Survey Point Source Catalogue contains 235,696,744 sources within $20<l<220$ and $-6<b<6$ (no selection cuts applied). These three master catalogues will form the basis of XGAPS.

\subsection{Proper motion corrections and cross-matching} \label{sec:matching}

To minimise mismatches between the \gaia\ DR3 and IGAPS, as well as recovering fast moving objects, it is important to take into account the proper motion of targets and reference epoch of all observations. \gaia\ DR3 provides proper motion for all systems satisfying the required quality cuts, and have astrometric measurements quoted to epoch J2016 for all sources. Because of the survey design, IGAPS does not provide proper motion information for targets, but does provide the epoch of observation for all targets individually.

Ideally, for precise cross-matching between the catalogues, the \gaia\ astrometry would have to be propagated to the IGAPS epoch of observation for each source individually before cross-matching is performed. This approach becomes unfeasible when considering large data tables. The approach used instead is similar to that used by \cite{gIPHAS}, but modified for IGAPS. The first step is to separate the merged IGAPS catalogue back into its IPHAS and UVEX constituents. This is because the epoch of observation is different between the two surveys. The next step is to separate the split IPHAS/UVEX catalogues into monthly batches based on the start of the $r$-band observation obtained for each individual IPHAS or UVEX detection separately. Because of the observing strategy of both IPHAS and UVEX, which sequentially observe all bands immediately following each other, we take the epoch of a particular target to be the start of the $r$-band observation as being representative of all other observations for that target. This ensures that the epoch-corrected positional uncertainty of the \gaia\ catalogue is relativity small even for high proper motion objects. For example, the recomputed \gaia\ coordinates for an object with an extreme proper motion of $2$''/year should be at worst $\approx 0.08$'' off the IGAPS position (if the epoch used was wrong by half a month).

Each corresponding IPHAS and UVEX monthly batch is then cross-matched with the master \gaia\ DR3 catalogue after having recomputed the \gaia\ astrometry to the mid-point epoch for each month. We then select the best positional closest \gaia\ DR3 match in the sky within a generous 1'' of a given IPHAS or UVEX entry. This results in a cross-match for each month and for each of the two IPHAS and UVEX surveys individually. Overall, 34,252,452 sources from IGAPS find a counterpart within \gaia\ DR3. These are split into 32,138,484 sources with detections in both IPHAS and UVEX, 1,562,330 sources with IPHAS-only detections, and 551,638 sources with UVEX-only detections. Inevitably there will be duplicated entries where multiple IPHAS or UVEX sources will have matched to the same \gaia\ DR3 source. These duplicate matches (265,272 of them) are removed by first concatenating all matched sources from the monthly batches for both IPHAS and UVEX together, and then performing an internal cross-match based solely on the unique \gaia\ DR3 source ID. Where multiple entries are encountered, preference is given, in order, to (i) sources that have both IPHAS and UVEX observation and (ii) have the smallest sky separation between the respective IPHAS/UVEX entry and \gaia\ DR3. At this stage the only duplicate entries present are those already flagged by \gaia\ DR3. These sources are retained, but can be easily removed at a later stage if required. The final number of sources in the XGAPS catalogue is 33,987,180. 

Having obtained a sub-arcsecond cross-match between IGAPS and \gaia\ DR3 the next step is to cross-match these with the UKIDSS GPS point source catalogue. A similar procedure is performed, where the UKIDSS data is first split into monthly batches based on the epoch of observation. These monthly batches are then cross-matched to the 33+ million sources based on the epoch corrected \gaia\ DR3 sky positions, resulting in 21,240,420 pairs. Duplicate UKIDSS matches (48 of them) are removed based on the \gaia\ DR3 source ID as previously done with the IGAPS cross-match, retaining the closest UKIDSS match to the corresponding \gaia\ source. Thus the total number of cross-matched UKIDSS sources is 21,240,381. It is important to note that although all sources in XGAPS will have \gaia\ DR3 information as well as either IPHAS or UVEX (or both), not all will necessarily have a UKIDSS counterpart.

The selection cuts described in Section \ref{sec:crossMatch} may introduce a number of mismatches between the IGAPS catalogues and the epoch-corrected \gaia\ catalogue. These miss-matches may arise due to crowding in the Galactic plane, and can be mostly attributed to the selection on \texttt{phot\_g\_mean\_flux\_over\_error}>3. A more detailed analysis of this effect has already been discussed in \cite{gIPHAS}. What was found is an upper limit of 0.1\% on the fraction of mismatches associated with their selection cut of \texttt{phot\_g\_mean\_flux\_over\_error}>5 in some of the most crowded regions of the Galactic plane mostly affecting the faintest sources. For XGAPS it is expected that the number of miss-matches is even lower than the 0.1\% miss-match fraction quoted in \cite{gIPHAS} as the selection cut now includes many more \gaia\ sources. The next section also introduces an additional quality flag that can be used to further clean erroneous matches and/or targets that have spurious astrometric solutions.

\section{Cleaning XGAPS with Random Forests} \label{sec:RF}
\begin{figure*}
    \includegraphics[width=1\columnwidth]{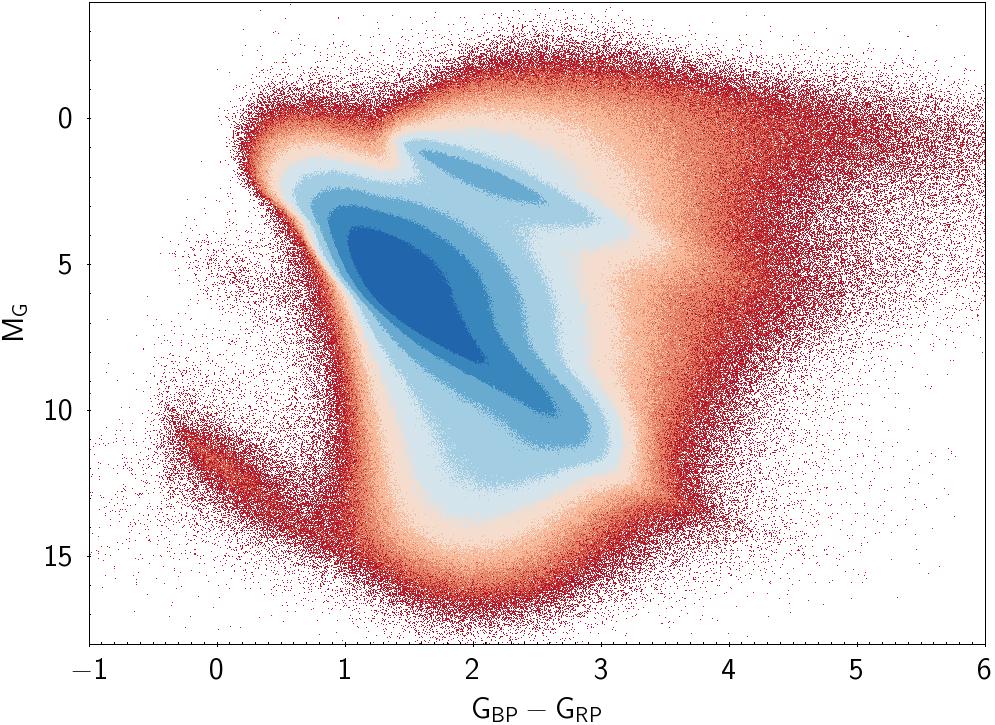}
    \includegraphics[width=1\columnwidth]{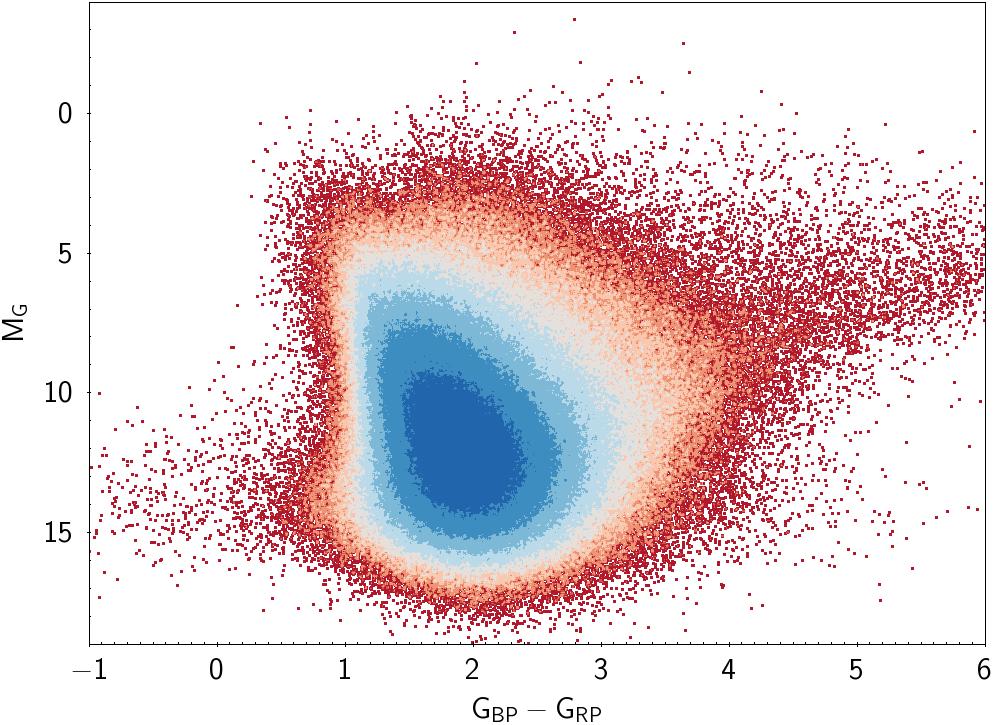}
    \caption{Left panel: \gaia-based absolute CMD of all cross-matched sources between \gaia\ and IGAPS. Right panel: The recovered negative parallax ``mirror sample'' from the \gaia/IGAPS cross-match. In producing this the absolute val;ue of the \gaia\ parallax measurements are used.}
    \label{fig:cmd1}
\end{figure*}

The left panel of Fig. \ref{fig:cmd1} shows colour-magnitude diagram (CMD) using the \gaia-based colours for all  cross-matched targets as described in Section \ref{sec:crossMatch}. The distances used to convert apparent to absolute magnitudes have been inferred via $M = m+5+5 \log_{10}(\varpi/1000)$, where $M$ and $m$ are the absolute and apparent magnitudes respectively, and $\varpi$ the parallax in milliarcseconds provided by \gaia\ DR3. \cite{gaiaEDR3Plx} provide a correction to the $\varpi$ measurements to correct for the zero point bias. This correction is not applied here, and neither is extinction, but users of XGAPS can do so through the available code provided by \cite{gaiaEDR3Plx}. 

As can be seen from the left panel of Fig. \ref{fig:cmd1} both CMDs appear to be ``polluted'' by spurious sources. This is particularly evident in the regions between the main sequence and white dwarf tracks, where a low population density of sources is expected. Similar contamination can also be observed in different colour combination CMD plots. Spurious astrometric solutions from \gaia\ can be due to a number of reasons. One of the major causes that produce such spurious parallax measurements is related to the inclusion of outliers in the measured positions. In \gaia\ DR3 this is more likely to occur in regions of high source density (as is the case in the Galactic plane) or for close binary systems (either real or due to sight line effects) which have not been accounted for. The dependence of spurious parallax measurements on other measured quantities in \gaia\ DR3 is not straight forward to disentangle, and CMDs cannot be easily cleaned through the use of empirical cuts on the available \gaia\ DR3 parameters.

Several methods attempting to identify spurious astrometric sources have been explored in the literature. \cite{gIPHAS} defined both a ``completeness'' and ``purity'' parameter that can be used to clean the resulting CMDs from the previous cross-match between \gaia\ DR2 and IPHAS. More recently, \cite{GCNS} employed a machine learning classifier based on Random Forests to identify spurious astrometric measurements in  the 100 pc sample of \gaia\ EDR3. In both cases, a negative parallax sample had been used to infer common properties of spurious astrometric sources. This was then generalised and applied to the positive parallax sources to identify spurious measurements. 

A classifier will only be as good at generalising a given set of properties as the provided training set allows. Here a Random Forest classifier is also used to clean XGAPS from the contamination of bad astrometric measurements. To explore this further, the same cross-matching method as described in Section \ref{sec:crossMatch} is performed using as a master \gaia\ catalogue of all sources satisfying the same quality cuts as described in Section \ref{sec:selection} but inverting the parallax signal-to-noise selection criteria to be less than $-3$ (\texttt{parallax\_over\_error}<-3). This produces a total of 1,034,661 sources after the cross-matching with the IGAPS catalogue has been performed. The right panel of Fig. \ref{fig:cmd1} shows the \gaia\ CMD of the recovered negative parallax ``mirror sample'' after having parsed through the same cross-matching pipeline as all other XGAPS sources. To obtain ``absolute magnitudes'' for sources the absolute value of the negative parallax has been used. It is clear from comparing both panels of Fig. \ref{fig:cmd1} that the suspiciously spurious parallax sources and negative parallax sources occupy similar regions of the CMDs. This in turn suggests that the same systematic measurement challenges are affecting both these samples, even though there is no clear parameter combination cut from the \gaia\ astrometric measurements that can be used to exclude spurious sources.

%\begin{figure}
%	\includegraphics[width=1\columnwidth]{Figures/CMDneg.jpeg}
%    \caption{The recovered negative parallax ``mirror sample'' from the \gaia/IGAPS cross-match.}
%    \label{fig:negCMD}
%\end{figure}

In a similar way to what has been adopted in \cite{GCNS} to remove spurious sources, a Random Forest (\citealt{RF}) is trained through the use of XGAPS data to classify all $\approx 34$ million entries into two categories (good vs. bad astrometric solutions) purely based on astrometric quantity and quality indicators provided by \gaia\ DR3 and augmented by astrometric indicators resulting from XGAPS. To achieve this a reliable training set of both categories is required. Because XGAPS sources are found in the crowded Galactic plane, and because these sources may suffer from specific systematic errors, a training/testing set is constructed from XGAPS data alone. The good astrometric solution set is compiled by selecting all sources in XGAPS which have a parallax signal-to-noise measurement above 5. This results in 19,242,307 good astrometric solution sources used for training. Although some bad parallax measurement sources may be expected to have a parallax signal-to-noise measurement above 5, it is reasonable to assume that a small fraction of sources will fall into this category. The bad astrometric training sources are compiled through the use of the ''negative parallax mirror sample'', for which the CMD is shown in the right panel of Fig. \ref{fig:cmd1}. This is obtained by selecting sources with a parallax signal-to-noise measurement below $-5$, resulting in 250,069 sources. In total, the set of good and bad astrometric solution targets is 19,492,876. The testing set is created by randomly selecting 20\% of the lowest populated class (50,113 from the bad astrometric sources), and randomly selecting the same number of sources from the other class. All remaining sources are used as a training set.

The classification model consists of a trained Random Forest (\citealt{RF}) using a total of 26 predictor variables listed in Table \ref{tab:RF} which are purely astrometry based. Each decision tree in the Random Forest is grown using 5 randomly chosen predictor variables, and each tree is grown to their full length. Surrogate splits when creating decision trees are used to take into account missing variables in some of the training samples. Each tree is grown by resampling targets, with replacement, in the training sample while keeping the total number of training samples per tree the same as the total number of targets used for training. Because the number of good astrometric training sources is much larger than the bad astrometric sources, each tree is grown using all bad astrometric sources (200,456 after having removed the testing set), and randomly under-sampling the same number of good training sources. This ensures that there is a balance between the two classes for each grown tree. These resampling techniques ensure that each tree is grown using a different subset of the training set and related predictors, which in turn avoids the Random Forest from overtraining (\citealt{RF}). In total, the Random Forest consists of 1001 decision trees. Final source classifications are assigned by the largest number of trees that classified the source as a particular class. The vote ratio between the two classes is also retained in the XGAPS catalogue. We have further attempted to establish the relative predictor importance for each of the 26 predictors used. This is achieved through the same classifier methodology described. However, for computational time purposes, the predictor importance values only are obtained by growing each tree using the same good training sources (200,456 randomly selected from the entire population) rather than resampling these for each individual tree. The resulting predictor importance using the out-of-bag samples during training is included in Table \ref{tab:RF}.

\begin{table}
\begin{center}
\begin{tabular}{l l}
Predictor Name & Predictor Importance\\
\hline\hline
    pmra                        &      11.68  \\
    pmdec                       &      9.07 \\
    bMJD\_separation\_UVEX        &      4.30 \\
    bMJD\_separation\_IPHAS       &      4.26 \\
    ipd\_frac\_multi\_peak         &      4.06 \\
    ipd\_gof\_harmonic\_amplitude  &      3.61 \\
    astrometric\_n\_good\_obs\_al   &      2.67 \\
    astrometric\_n\_obs\_al        &      2.65 \\
    scan\_direction\_mean\_k1      &      2.53 \\
    parallax\_error              &      2.42 \\
    scan\_direction\_mean\_k2      &      2.24 \\
    scan\_direction\_mean\_k3      &      2.22 \\
    ruwe                        &      1.96 \\
    astrometric\_excess\_noise\_sig &      1.84 \\
    astrometric\_gof\_al          &      1.81 \\
    astrometric\_excess\_noise    &      1.74 \\
    pmdec\_error                 &      1.70 \\
    redChi2                     &      1.64 \\
    scan\_direction\_strength\_k1  &      1.57 \\
    astrometric\_sigma5d\_max     &      1.50 \\
    ipd\_frac\_odd\_win            &      1.49 \\
    scan\_direction\_mean\_k4      &      1.49 \\
    astrometric\_n\_bad\_obs\_al    &      1.42 \\
    astrometric\_chi2\_al         &      1.36 \\
    pmra\_error                  &      1.33 \\
    astrometric\_n\_obs\_ac        &      0.27 \\ 
    \hline
\end{tabular}
\caption{Out-of-bag predictor importance of all predictors used for classification by the Random Forest classifier ordered according to importance. The predictor names used in the table correspond to column names used in the XGAPS catalogue. A short description of each can be found in the Appendix.
}
\end{center}
\label{tab:RF} 
\end{table}

\begin{figure}
	\includegraphics[width=1\columnwidth]{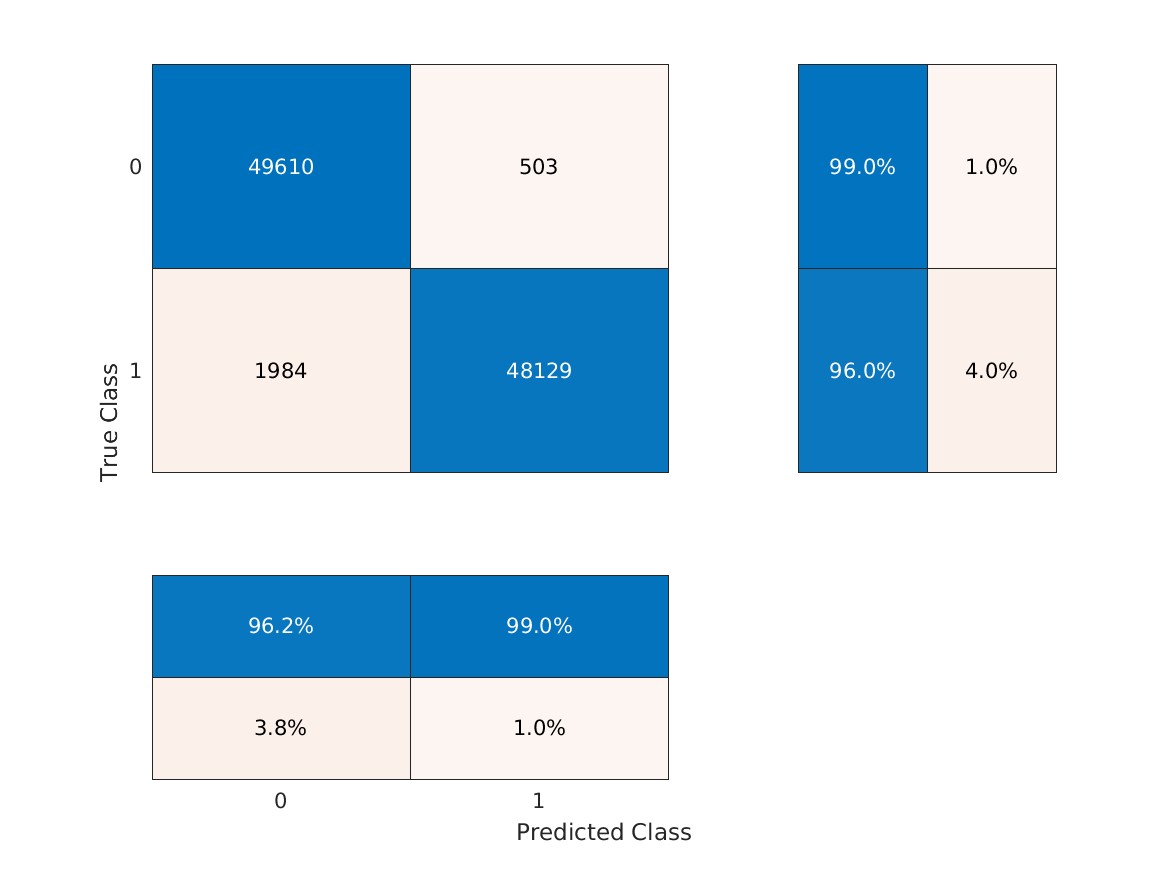}
    \caption{Confusion matrix between the positive and negative parallax samples computed on the test set. Class values of 0 represent "bad" astrometric sources while a value of 1 represent "good" astrometric sources. Details of the definition of the test set and training of the Random Forest can be found in Section \ref{sec:RF}.}
    \label{fig:conf}
\end{figure}

The Random Forest is robust against variations in the number of trees or candidate predictors, as altering these did not produce substantially different results as evaluated on the test set. It is important to note that although the bad training sources can be considered to be the result of bonafide spurious astrometric measurements, some systems in the good training set are expected to have been mislabeled by the training set selection criteria. Thus when inspecting the Random Forest classification accuracy on the testing set only sources with misclassified labels from the bad astrometric sources should be considered, and these should provide a lower limit on the true accuracy of the classifier. The final result on the testing set is summarised by the confusion matrix shown in Figure \ref{fig:conf}. Overall 1984 sources are classified as bad sources owning a parallax signal-to-noise measurement above 5. More importantly, 503 out of 50,113 bad astrometric sources (1.0\%) have been mislabeled, and these should provide the lower limit on the accuracy of the classifier.

\begin{figure}
	\includegraphics[width=1\columnwidth]{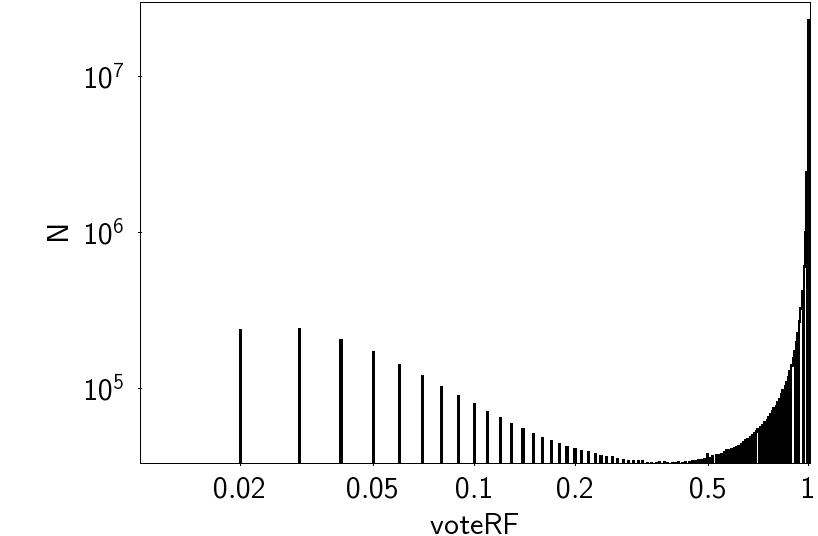}
    \caption{Distribution of associated votes as computed by the trained Random Forest described in Section \ref{sec:RF} for the full XGAPS targets. The \texttt{voteRF} value is included in the XGAPS catalogue for each source. Objects with \texttt{voteRF}>0.5 have a \texttt{flagRF} value of 1 in XGAPS rather than 0.} 
    \label{fig:voteRF}
\end{figure}

Having trained the classification model, all $\approx 34$ million sources in XGAPS are parsed through the Random Forest classifier and receive an associated vote (see Fig. \ref{fig:voteRF}) from each tree and an associated flag with the predicted classification. Sources are classified as good astrometric sources if more than 50\% of individual trees in the Random Forest classifier have classified them as such, and are assigned a flag in the catalogue of \texttt{flagRF}=1. If this is not achieved, the source flags are set to \texttt{flagRF}=0. This results in 30,927,929 (91\%) targets with \texttt{flagRF}=1 and 3,059,251 (9\%) with \texttt{flagRF}=0.

%It is important to recall that all predictor variables are purely astrometric based, and no colour or magnitude information is used in the training process. 

\begin{figure*}
	\includegraphics[width=1\columnwidth]{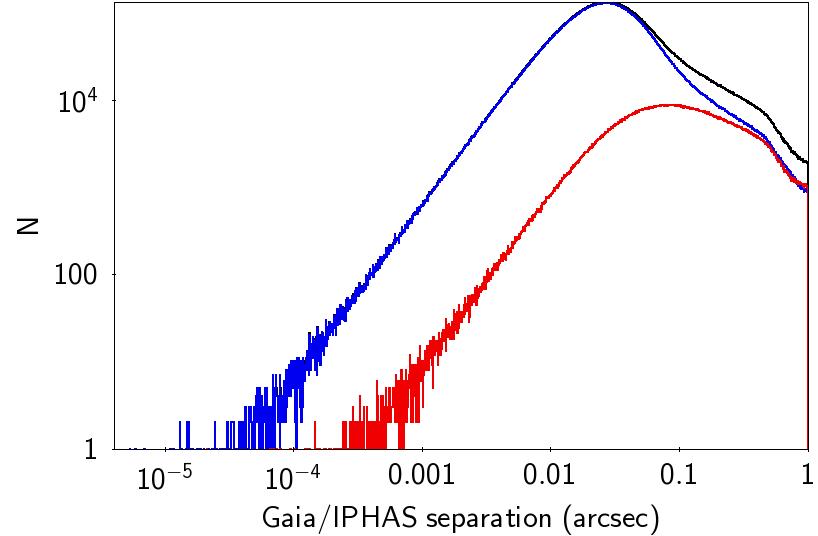}
    \includegraphics[width=1\columnwidth]{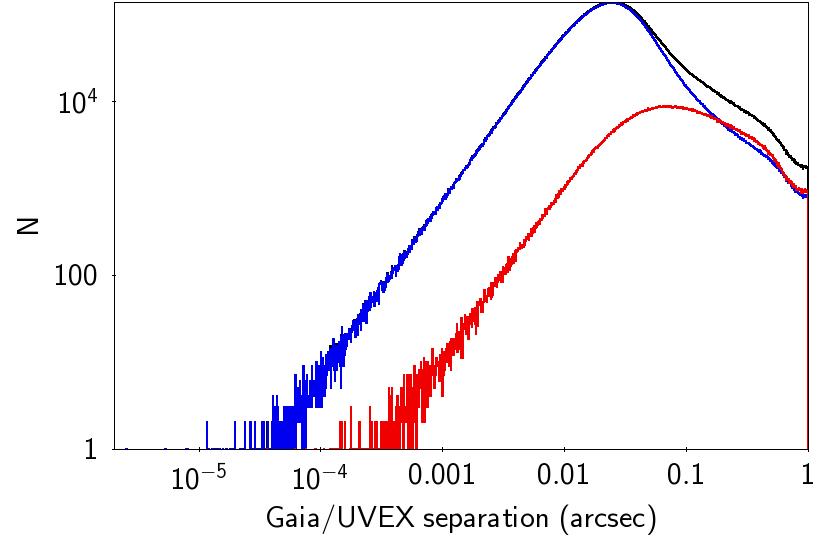}
    \caption{Distribution of the separations between all matched sources in XGAPS between the \gaia/IPHAS targets (left) and \gaia/UVEX targets (right) are shown with black solid lines. Both panels also show the decomposition of the distribution employing the Random Forest classifier to select ``good'' astrometric targets (\texttt{flagRF}=1, blue solid lines) and``bad'' astrometric targets (\texttt{flagRF}=0, red solid lines).}
    \label{fig:separation}
\end{figure*}

Fig. \ref{fig:separation} shows the angular separation between the individual IPHAS and UVEX matches to the epoch-corrected \gaia\ DR3 sources. The bulk of the population finds an angular separation of about 0.02 arcseconds, but there exists an additional component of sources evident at larger separations. Although these sources have found their correct match between \gaia\ and both IPHAS and UVEX, the larger angular separation may in fact be attributed to poor astrometry in \gaia\ DR3. Shown in the same figure are also the distribution of the good vs. bad astrometric sources as classified using the trained Random Forest. It is clear that the classifier has been able to separate those sources with relatively large angular separation when compared to the bulk of the population. 

\begin{figure*}
	\includegraphics[width=1\columnwidth]{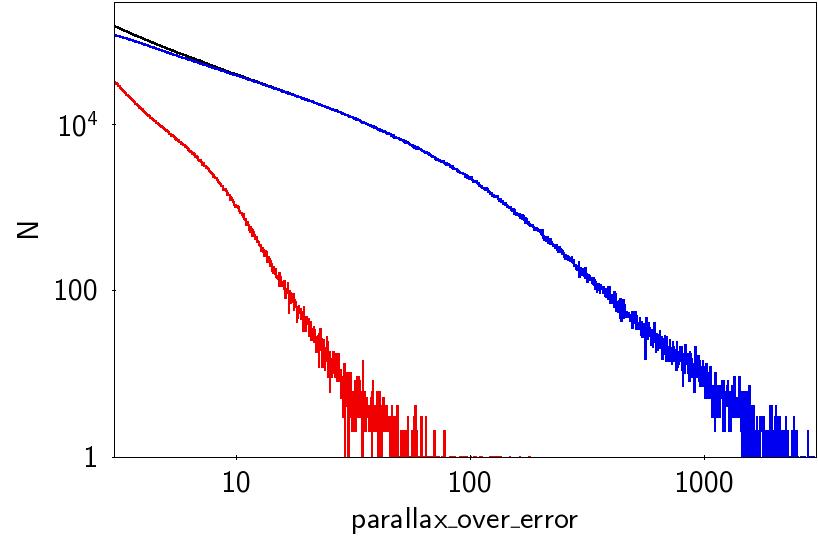}
    \includegraphics[width=1\columnwidth]{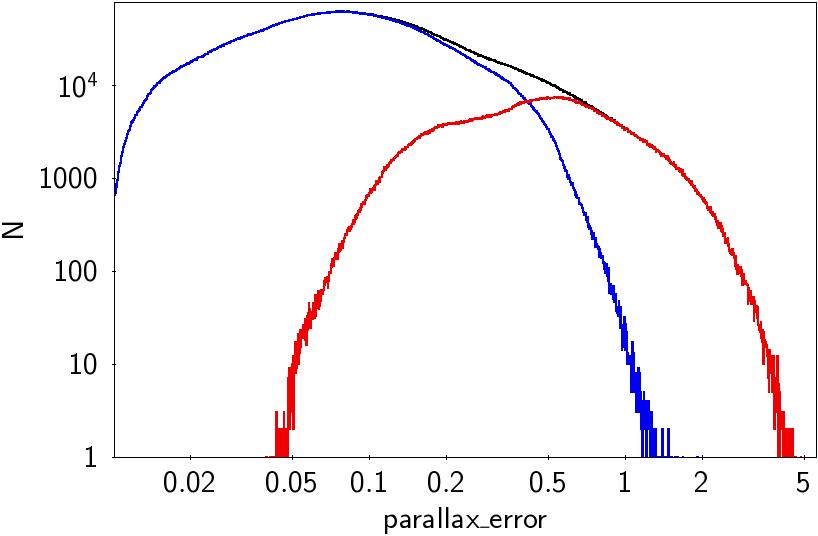}
    \newline
    \includegraphics[width=1\columnwidth]{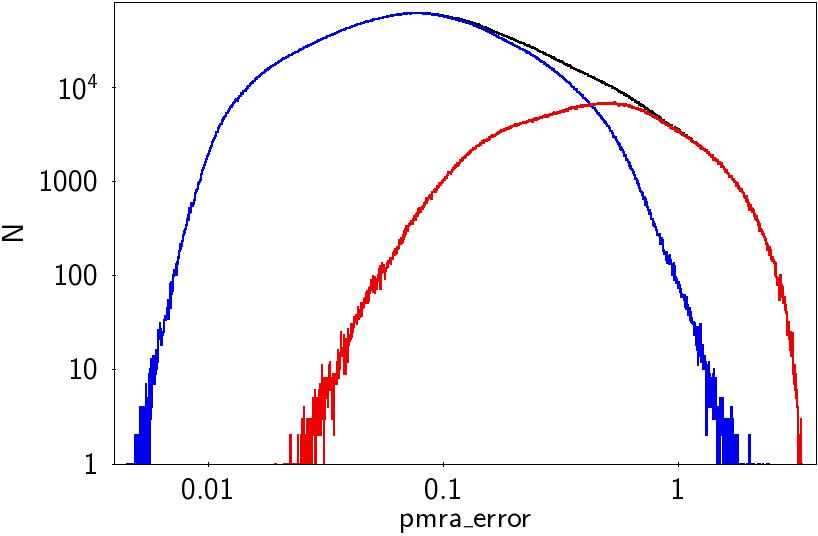}
    \includegraphics[width=1\columnwidth]{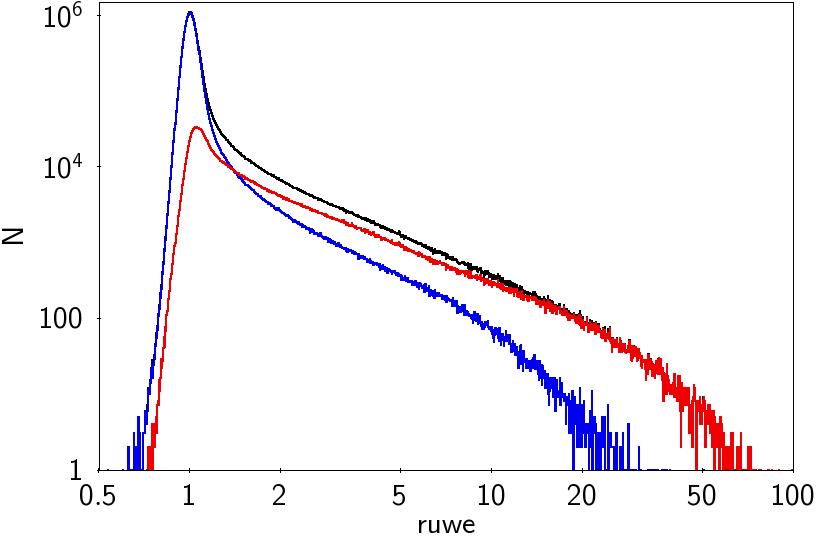}
    \caption{Distributions of a subset of astrometric parameters taken from \gaia\ DR3 included in XGAPS (black solid lines). All but the \texttt{parallax\_over\_error} values have been used for training the Random Forest classifier. All panels also show the decomposition of the distribution employing the Random Forest classifier to select ``good'' astrometric targets (\texttt{flagRF}=1, blue solid lines) and ``bad'' astrometric targets (\texttt{flagRF}=0, red solid lines).}
    \label{fig:predVars}
\end{figure*}

This split between the good vs. bad astrometric sources can also be validated when considering other astrometric predictor variables used by the classifier. Fig. \ref{fig:predVars} shows the distributions of an additional 3 predictor variables (\texttt{parallax\_error}, \texttt{pmra\_error}, \texttt{ruwe}) as well as the parallax signal-to-noise measurement (\texttt{parallax\_over\_error}) which has been used to select the training set. In all cases the Random Forest classifier appears to have separated the apparent bimodal distributions observed in the predictor variables. 

\begin{figure*}
	\includegraphics[width=1\columnwidth]{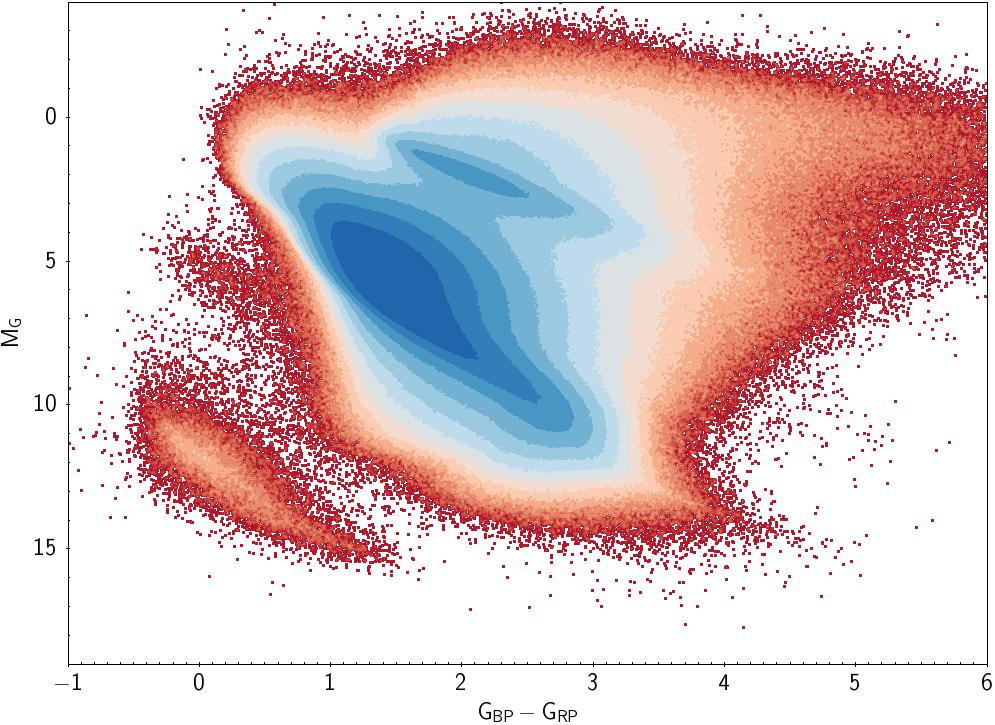}
    \includegraphics[width=1\columnwidth]{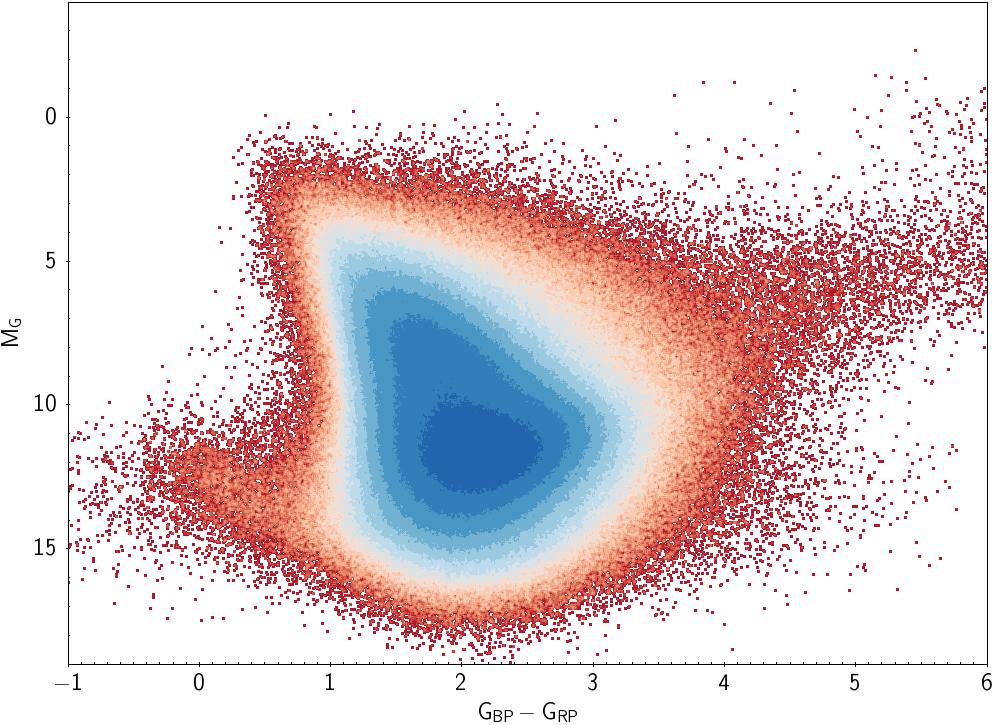}
    \caption{\gaia-based absolute CMDs for all targets in the XGAPS catalogue. The panel on the left shows all targets with \texttt{flagRF}=1, while targets with \texttt{flagRF}=0 are displayed in the right panel. Although all sources displayed have a positive parallax measurement, the "bad" astrometric sample in XGAPS as defined by the Random Forest occupies a similar region in CMD space as the negative parallax ``mirror sample'' used for training and shown in the right panel Fig. \ref{fig:cmd1}.}
    \label{fig:gaiaRF}
\end{figure*}

Inspecting the CMDs of the predicted good vs. bad astrometric targets provides additional insight on the Random Forest performance. Fig. \ref{fig:gaiaRF} displays the \gaia\ CMD of the predicted good vs. bad astrometric targets. Overall, the Random Forest classifies a total of 30,944,717 good astrometric targets ($\approx$91\%) and 3,042,463 bad astrometric targets ($\approx$9\%). It is clear that most of the bad astrometric sources are correctly removed as they populate the same region in the CMD as the negative parallax sample used for training (see right panel of Fig. \ref{fig:cmd1}). Although the split has been efficiently achieved, it is also the case that some good astrometric sources have been flagged as bad ones by the classifier, and vice-versa. This is particularly evident when inspecting the CMD region for sources classified as having good astrometry (left panel in Fig. \ref{fig:gaiaRF}), which appears to still be populated with relatively large number of sources on the blue side of the main sequence. Furthermore, some sources flagged as having bad astrometry by the classifier appear to populate the WD track, and it is also possible some of these have been mislabeled (right panel in Fig. \ref{fig:gaiaRF}). Overall however, the bulk of the bad astrometric sources appears to have been removed correctly. 
 
\section{Potential applications of the XGAPS catalogue} \label{sec:catalogue}

Owning broad and narrow-band photometric measurements for $\approx 34$ million Galactic plane sources, astrometric information, as well as multi-epoch photometry in many of these, the applications for the XGAPS catalogue can be wide-reaching, especially for the identification of specific source types and related population studies. Examples based on the \gaia/IPHAS catalogue (\citealt{gIPHAS}) include the discovery of new binary systems (\citealt{carrell22}), the selection and identification of Herbig Ae/Be systems (\citealt{vioque20}), planetary nebulae (\citealt{sabin22}), as well as candidate X-ray emitting binaries (\citealt{gandhi21}). Further applications may also be found in constructing reliable training sets for classification, as has been used by \cite{gaiaSynth} to train a Random Forest for classification of targets based on synthetic photometry.

\begin{figure*}
	\includegraphics[width=1\columnwidth]{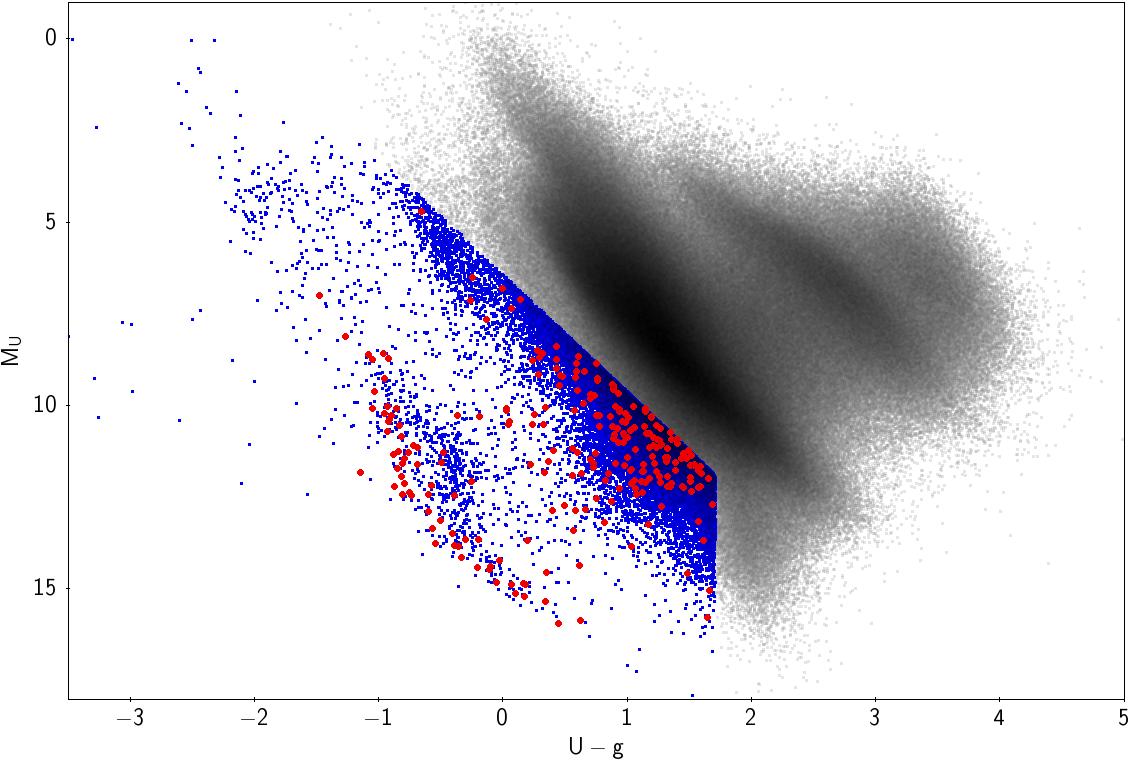}
    \includegraphics[width=1\columnwidth]{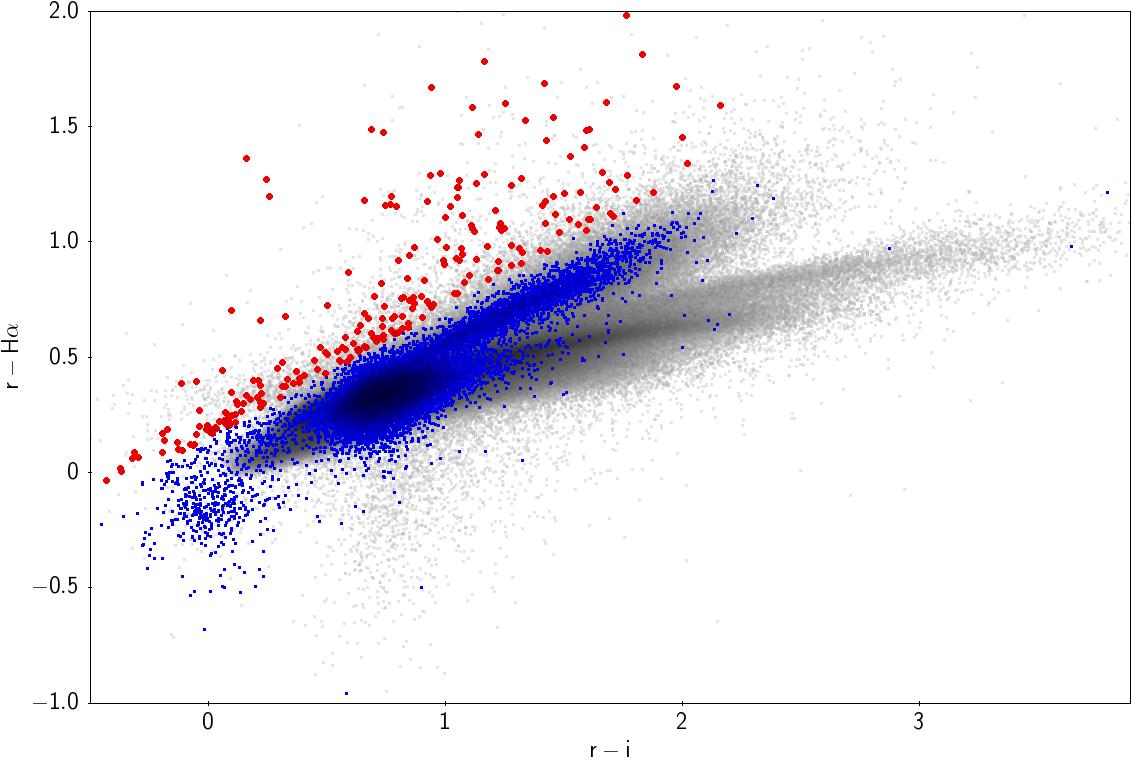}
    \caption{\gaia/UVEX CMD (left panel) and corresponding IPHAS-based colour-colour diagrams demonstrating simple selection cuts to select candidate white dwarf systems to be observed by 4MOST (\citealt{4most}). Gray points in both panels show all targets in XGAPS with declination smaller than 5 degrees (observable from Paranal) and \texttt{flagRF}=1. Blue points mark targets selected as blue-excess sources, likely related to white dwarf emission contributing to the UVEX photometry. The red points mark blue-excess candidates that also display evidence of \ha-excess emission as determined from the IPHAS photometry. The exact cuts are described in Section \ref{sec:catalogue}.}
    \label{fig:wdbSelection}
\end{figure*}

Also important, XGAPS provides information that can be efficiently used in selecting targets for large multi-object spectroscopic surveys such as the WHT Enhanced Area Velocity Explorer (WEAVE: \citealt{weave}) and the 4-metre Multi-Object Spectrograph Telescope (4MOST: \citealt{4most}). An example of this is the selection of white dwarf candidates in the Galactic plane to be observed with 4MOST as part of the community selected White Dwarf Binary Survey (PIs: Toloza and Rebassa-Mansergas). This includes a total of 28,102 targets that satisfy the following criteria in XGAPS:
\begin{itemize}
\item Have a \gaia\ declination $<5$ degrees
\item Have the \texttt{flagRF} set to 1
\item Lie within the region $M_{U} > 3.20 \times (U-g) + 6.42$ and $(U-g)<1.71$
\end{itemize}
\noindent The resulting CMD using the UVEX colours is shown in the left panel of Fig. \ref{fig:wdbSelection}. The declination cut was employed to ensure targets are observable from Paranal Observatory where the 4MOST survey will be carried out from. The \texttt{flagRF} is employed to minimise spurious cross-matches and bad astrometric targets. The final colour-magnitude cuts are somewhat ad-hoc at this stage (especially as the $U_{RGO}$ band has not yet been photometrically calibrated across the full survey), but attempt to select all blue-excess sources relative to the main sequence as defined in the UVEX passbands (the bluest set of the XGAPS catalogue). Although preliminary and in need of refinement using well-validated and spectroscopically confirmed targets, these colour cuts provide a first attempt to select white dwarf candidates in the plane for the 4MOST survey. A further cut using the IPHAS passbands of $(r-H\alpha)>0.56 \times (r-i) + 0.27$ to select \ha-excess sources yields 241 likely accreting white dwarf systems (right panel of Fig. \ref{fig:wdbSelection}). We point out that these colour cuts are preliminary, and only serve to demonstrate the potential application of the XGAPS catalogue. Specifically for the selection of \ha-excess sources, a more refined method of selecting \ha-excess candidates based on the local population as defined in absolute colour-magnitude diagrams has been shown to produce more complete samples of objects, but this comes at the expense of purity (e.g. \citealt{fratta21}).

\section{Conclusion} \label{sec:conclusion}

We have presented the XGAPS catalogue which provides a sub-arcsecond cross-match between \gaia\ DR3, IPHAS, UVEX and UKIDSS. It contains photometric and astrometric measurements for $\approx 34$ million sources within the northern Galactic plane. In total, XGAPS contains 2 epoch photometry in the $r$-band, as well as single-epoch (not simultaneous) photometry in up to 9 broad-band filters ($U_{RGO}$, $g$, $r$, $i$, $J$, $H$, $K$, $G$, $G_{BP}$ and $G_{RP}$) and one narrow-band \ha-filter. XGAPS additionally provides a confidence metric inferred using Random Forests aimed at assessing the reliability of the \gaia\, astrometric parameters for any given source in the catalogue. XGAPS is provided as a catalogue with 111 columns. A description of the columns is presented in Table \ref{tab:cols}. The full XGAPS catalogue can be obtained through ViZieR. As XGAPS only covers the northern Galactic plane, future extensions are planned to merge the southern Galactic plane and bulge using data from the VST Photometric \ha\ Survey of the Southern Galactic Plane and Bulge (VPHAS+: \citealt{vphas}).

\section*{Acknowledgements}
Cross-matching between catalogues has been performed using STILTS, and diagrams were produced using the astronomy-oriented data handling and visualisation software TOPCAT (\citealt{topcat}). This work has made use of the Astronomy \& Astrophysics package for Matlab (\citealt{matlabOfek}). This research has also made extensive use of the SIMBAD database, operated at CDS, Strasbourg. This work is based on observations made with the Isaac Newton Telescope operated on the island of La Palma by the Isaac Newton Group of Telescopes in the Spanish Observatorio del Roque de los Muchachos of the Instituto de Astrofísica de Canarias. This work is also based in part on data obtained as part of the UKIRT Infrared Deep Sky Survey. This work has made additional use of data from the European Space Agency (ESA) mission {\it Gaia} (\url{https://www.cosmos.esa.int/gaia}), processed by the {\it Gaia} Data Processing and Analysis Consortium (DPAC, \url{https://www.cosmos.esa.int/web/gaia/dpac/consortium}). Funding for the DPAC has been provided by national institutions, in particular the institutions participating in the {\it Gaia} Multilateral Agreement.

MM work was funded by the Spanish MICIN/AEI/10.13039/501100011033 and by ``ERDF A way of making Europe'' by the ``European Union'' through grant RTI2018-095076-B-C21, and the Institute of Cosmos Sciences University of Barcelona (ICCUB, Unidad de Excelencia ``Mar\'{\i}a de Maeztu'') through grant CEX2019-000918-M. ARM acknowledges support from Grant RYC-2016-20254 funded by MCIN/AEI/10.13039/501100011033 and by ESF Investing in your future, and from MINECO under the PID2020-117252GB-I00 grant.

%%%%%%%%%%%%%%%%%%%%%%%%%%%%%%%%%%%%%%%%%%%%%%%%%%
\section*{Data Availability}
The XGAPS catalogue produced in this paper is available and can be found on VizieR.
%%%%%%%%%%%%%%%%%%%%%%%%%%%%%%%%%%%%%%%%%%%%%%%%%


\bibliographystyle{mnras}
\bibliography{XGAPS} % if your bibtex file is called example.bib
%%%%%%%%%%%%%%%%%%%%%%%%%%%%%%%%%%%%%%%%%%%%%%%%%%

%%%%%%%%%%%%%%%%% APPENDICES %%%%%%%%%%%%%%%%%%%%%
%%% TABLE %%%
\newpage
\appendix

\onecolumn
\section{Catalogue format}
\small
\begin{longtable}{p{4cm}p{1.5cm}p{10cm}}
\caption{\label{tab:cols} Definition of columns in the XGAPS catalogue. In total the catalogue contains 111 columns.}\\

\hline
Column name & Unit & Description \\
\hline
\endfirsthead

\multicolumn{3}{c}%
{{\bfseries \tablename\ \thetable{} -- continued}} \\
\hline
Column name & Unit & Description \\ 
\hline
\endhead

\hline \hline
\endlastfoot
\texttt{GaiaDR3} & & Unique \textit{Gaia} DR3 source designation \\
\texttt{RAJ2016\_Gaia} & degrees & \textit{Gaia} DR3 barycentric right ascension (ICRS) at Epoch 2016.0 \\
\texttt{DEJ2016\_Gaia} & degrees & \textit{Gaia} DR3 barycentric declination (ICRS) at Epoch 2016.0 \\
\texttt{err\_RAJ2016\_Gaia} & mas & Standard error of right ascension (err\_RAJ2016\_Gaia$\times$cos(dec)) \\
\texttt{err\_DEJ2016\_Gaia} & mas & Standard error of declination \\
\texttt{parallax} & mas & Absolute stellar parallax\\
\texttt{parallax\_error} & mas & Standard error of parallax\\
\texttt{pm} & mas/yr & Total proper motion \\
\texttt{pmra} & mas/yr & Proper motion in right ascension direction (\texttt{pmra}$\times$cos(\texttt{DEJ2016\_Gaia})) \\
\texttt{pmdec} & mas/yr &  Proper motion in declination direction\\
\texttt{pmra\_error} & mas/yr & Standard error of proper motion in right ascension direction\\
\texttt{pmdec\_error} & mas/yr & Standard error of proper motion in declination direction\\
\texttt{astrometric\_excess\_noise} & mas & Excess noise of the source from the \textit{Gaia} astrometric solution\\
\texttt{astrometric\_excess\_noise\_sig} &  & Significance of excess noise of the source from the \textit{Gaia} astrometric solution\\
\texttt{ruwe} & & Renormalised unit weight error \\
\texttt{duplicated\_source} & & Source with multiple source identifiers in \textit{Gaia DR3}\\
\texttt{phot\_g\_mean\_flux} & electron/s & \textit{Gaia} DR3 G-band mean flux\\
\texttt{phot\_g\_mean\_flux\_error} & electron/s &  Error on the G-band mean flux\\
\texttt{phot\_g\_mean\_flux\_over\_error} &  &  Integrated mean G flux divided by its error.\\
\texttt{phot\_bp\_mean\_flux} & electron/s & \textit{Gaia} DR3 integrated BP mean flux\\
\texttt{phot\_bp\_mean\_flux\_error} & electron/s & Error on the integrated BP mean flux\\
\texttt{phot\_bp\_mean\_flux\_over\_error} &  &  Integrated mean BP flux divided by its error\\
\texttt{phot\_rp\_mean\_flux} & electron/s & \textit{Gaia} DR3 integrated RP mean flux\\
\texttt{phot\_rp\_mean\_flux\_error} & electron/s & Error on the integrated RP mean flux\\
\texttt{hot\_rp\_mean\_flux\_over\_error} &  &  Integrated mean RP flux divided by its error\\
\texttt{nameIPHAS} & & Source designation (JHHMMSS.ss+DDMMSS.s) without IGAPS prefix for IPHAS detection\\
\texttt{RAJbMJD\_IPHAS} & deg & J2000 RA (Gaia DR2 reference frame) for IPHAS detection\\
\texttt{DECJbMJD\_IPHAS} & deg & J2000 DEC (Gaia DR2 reference frame) for IPHAS detection\\
\texttt{sourceID\_IPHAS} & & IPHAS Unique source identification string (run-ccd-detection number)\\
\texttt{posErr\_IPHAS} & arcsec & Astrometric fit error (RMS) across the CCD for IPHAS detection.\\
\texttt{rMJD\_I} & & Modified Julian Date at the start of the r\_I exposure.\\
\texttt{bMJD\_IPHAS} & d & Modified Julian Date used for cross-matching \textit{Gaia} to IPHAS\\
\texttt{bMJD\_separation\_IPHAS} & arcsec & Angular separation between the rewound \textit{Gaia} position at Epoch \texttt{bMJD\_IPHAS} to the nominal IPHAS position\\
\texttt{nameUVEX} & & Source designation (JHHMMSS.ss+DDMMSS.s) without IGAPS prefix for UVEX detection.\\
\texttt{RAJbMJD\_UVEX} & deg & J2000 RA (Gaia DR2 reference frame) for UVEX detection.\\
\texttt{DECJbMJD\_UVEX} & deg & J2000 DEC (Gaia DR2 reference frame) for UVEX detection.\\
\texttt{sourceID\_UVEX} & & Unique source identification string (run-ccd-detection number) for UVEX detection.\\
\texttt{posErr\_UVEX} & arcsec & Astrometric fit error (RMS) across the CCD for UVEX detection.\\
\texttt{rMJD\_U} & & Modified Julian Date at the start of the r\_U exposure.\\
\texttt{bMJD\_UVEX} & d & Modified Julian Date used for cross-matching \textit{Gaia} to UVEX\\
\texttt{bMJD\_separation\_UVEX} & arcsec & Angular separation between the rewound \textit{Gaia} position at Epoch \texttt{bMJD\_UVEX} to the nominal UVEX position\\
\texttt{sourceID\_UKIDSS} & & Unique UKIDSS identifier\\
\texttt{RAJbMJD\_UKIDSS} & deg & UKIDSS detection RA\\
\texttt{DECJbMJD\_UKIDSS} & deg & UKIDSS detection DEC\\
\texttt{epoch\_UKIDSS} & yr & Epoch at the start of the UKIDSS observation\\
\texttt{bMJD\_UKIDSS} & d & Modified Julian Date used for cross-matching \textit{Gaia} to UKIDSS\\
\texttt{bMJD\_separation\_UKIDSS} & arcsec & Angular separation between the rewound \textit{Gaia} position at Epoch \texttt{bMJD\_UKIDSS} to the nominal UKIDSS position\\
\texttt{phot\_g\_mean\_mag} & mag & Integrated G-band mean magnitude\\
\texttt{phot\_bp\_mean\_mag} & mag & Integrated BP mean magnitude\\
\texttt{phot\_rp\_mean\_mag} & mag & Integrated RP mean magnitude\\
\texttt{i} & mag & IPHAS i mag (Vega) using the 2.3 arcsec aperture.\\
\texttt{iErr} & mag & IPHAS i mag (Vega) error using the 2.3 arcsec aperture.\\
\texttt{ha} & mag & IPHAS \ha\ mag (Vega) using the 2.3 arcsec aperture.\\
\texttt{haErr} & mag & IPHAS \ha\ mag (Vega) error using the 2.3 arcsec aperture.\\
\texttt{r\_I} & mag & IPHAS r mag (Vega) using the 2.3 arcsec aperture.\\
\texttt{rErr\_I} & mag & IPHAS r mag (Vega) error using the 2.3 arcsec aperture.\\
\texttt{i2} & mag & IPHAS i mag (Vega) for the secondary detection.\\
\texttt{i2Err} & mag & IPHAS i mag (Vega) error for the secondary detection.\\
\texttt{ha2} & mag & IPHAS \ha\ mag (Vega) for the secondary detection.\\
\texttt{ha2Err} & mag & IPHAS \ha\ mag (Vega) error for the secondary detection.\\
\texttt{r2\_I} & mag & IPHAS r mag (Vega) for the secondary detection.\\
\texttt{r2Err\_I} & mag & IPHAS r mag (Vega) error for the secondary detection.\\
\texttt{r2MJD\_I} & d & Modified Julian Date at the start of the r2\_I exposure.\\
\texttt{r\_U} & mag & UVEX r mag (Vega) using the 2.3 arcsec aperture.\\
\texttt{rErr\_U} & mag & UVEX r mag (Vega) error using the 2.3 arcsec aperture.\\
\texttt{g} & mag & UVEX g mag (Vega) using the 2.3 arcsec aperture.\\
\texttt{gErr} & mag & UVEX g mag (Vega) error using the 2.3 arcsec aperture.\\
\texttt{U\_RGO} & mag & UVEX U$_{RGO}$ mag (Vega) using the 2.3 arcsec aperture.\\
\texttt{UErr} & mag & Random uncertainty for U\_RGO. Pipeline random error only\\
\texttt{r2\_U} & mag & UVEX r mag (Vega) for the secondary detection.\\
\texttt{r2Err\_U} & mag & UVEX r mag (Vega) error for the secondary detection.\\
\texttt{g2} & mag & UVEX g mag (Vega) for the secondary detection.\\
\texttt{g2Err} & mag & UVEX g mag (Vega) error for the secondary detection. \\
\texttt{U\_RGO2} & mag & UVEX U$_{RGO}$ mag (Vega) for the secondary detection.\\
\texttt{U2Err} & mag & UVEX Random uncertainty for U\_RGO2. Pipeline random error only\\
\texttt{r2MJD\_U} & d & Modified Julian Date at the start of the r2\_U exposure.\\
\texttt{j} & mag & UKIDSS j mag (Vega) using 2.0 arcsec aperture.\\
\texttt{jErr} & mag & UKIDSS j mag (Vega) error using 2.0 arcsec aperture.\\
\texttt{h} & mag & UKIDSS h mag (Vega) using 2.0 arcsec aperture.\\
\texttt{hErr} & mag & UKIDSS h mag (Vega) error using 2.0 arcsec aperture.\\
\texttt{k} & mag & UKIDSS k mag (Vega) using 2.0 arcsec aperture.\\
\texttt{kErr} & mag & UKIDSS k mag (Vega) error using 2.0 arcsec aperture. \\
\texttt{bp\_rp} & mag & BP - RP colour\\
\texttt{bp\_g} & mag & BP - G colour\\
\texttt{g\_rp} & mag & G - RP colour\\
\texttt{rmi} & mag & r\_I - i colour\\
\texttt{rmha} & mag & r\_I - \ha\ colour\\
\texttt{Umg} & mag & U\_RGO - g colour\\
\texttt{gmr} & mag & g - r\_U colour\\
\texttt{jmh} & mag & j - h colour\\
\texttt{jmk} & mag & j - k colour\\
\texttt{Gmj} & mag & G - j colour\\
\texttt{Umk} & mag & U\_RGO - k colour\\
\texttt{paramsSolved} & & Number of parameters solved for in the \textit{Gaia} DR3 model\\
\texttt{redChi2} & & Reduced chi2 for the \textit{Gaia} DR3 astrometric fit computed as \texttt{astrometric\_chi2\_al}/(\texttt{astrometric\_n\_good\_obs\_al}-\texttt{paramsSolved})\\
\texttt{rmsG} & & Root mean square for the G band \textit{Gaia} observations computed as \texttt{phot\_g\_mean\_flux\_error}*(sqrt(\texttt{phot\_g\_n\_obs}))\\
\texttt{frac\_rmsG} & & Fractional root mean square for the G band \textit{Gaia} observations computed as \texttt{phot\_g\_mean\_flux\_error}*(sqrt(\texttt{phot\_g\_n\_obs})/\texttt{phot\_g\_mean\_flux})\\
\texttt{dist} & pc & Inverse parallax distance to the source (no reddening) computed as 1/(abs(\texttt{parallax})/1000)\\
\texttt{pmT} & mas/yr & Transverse proper motion computed as sqrt(pow(\texttt{pmra},2)+pow(\texttt{pmdec},2))\\
\texttt{vT} & km/s & Transverse velocity computed as 4.74*\texttt{dist}*(\texttt{pm}/1000)\\
\texttt{M\_G} & mag & Absolute \textit{Gaia} G magnitude inferred using the inverse parallax distance \texttt{dist}\\
\texttt{R\_IPHAS} & mag & Absolute IPHAS r magnitude inferred using the inverse parallax distance \texttt{dist}\\
\texttt{I\_IPHAS} & mag & Absolute IPHAS i magnitude inferred using the inverse parallax distance \texttt{dist}\\
\texttt{R\_UVEX} & mag & Absolute UVEX r magnitude inferred using the inverse parallax distance \texttt{dist}\\
\texttt{G\_UVEX} & mag & Absolute UVEX g magnitude inferred using the inverse parallax distance \texttt{dist}\\
\texttt{U\_UVEX} & mag & Absolute UVEX U\_RGO magnitude inferred using the inverse parallax distance \texttt{dist}\\
\texttt{M\_j} & mag & Absolute UKIDSS j magnitude inferred using the inverse parallax distance \texttt{dist}\\
\texttt{M\_h} & mag & Absolute UKIDSS h magnitude inferred using the inverse parallax distance \texttt{dist}\\
\texttt{M\_h} & mag & Absolute UKIDSS k magnitude inferred using the inverse parallax distance \texttt{dist}\\
\texttt{voteRF} & & Random Forest classification probability for the source being classed as a good astrometric source\\
\texttt{flagRF} & & Random Forest classification. \texttt{flagRF}=1 if \texttt{voteRF}>0.5, else \texttt{flagRF}=0\\
\end{longtable}
\normalsize
\twocolumn

%%%%%%%%%%%%%%%%%%%%%%%%%%%%%%%%%%%%%%%%%%%%%%%%%%

% Don't change these lines
\bsp	% typesetting comment
\label{lastpage}
\end{document}